\documentclass[aps,prl,superscriptaddress,twocolumn,showpacs,preprintnumbers,amsmath,amssymb]{revtex4-1}
\usepackage{graphicx}
\usepackage{dcolumn}
\usepackage{bm}
\usepackage{upgreek}
\usepackage{placeins}

\begin{document}

\title{Disparity of superconducting and pseudogap scales in low-$T_\text{c}$ Bi-2201 cuprates}

\author{Th. Jacobs}
\author{S. O. Katterwe}
\author{H. Motzkau}
\author{A. Rydh}
\affiliation{Department of Physics, Stockholm University, AlbaNova University Center, SE-10691 Stockholm, Sweden}

\author{A. Maljuk}
\affiliation{Leibniz Institute for Solid State and Materials Research IFW Dresden, Helmholtzstr. 20, D-01171 Dresden, Germany}

\author{\hbox{T. Helm}}
\affiliation{Walther-Meissner-Institut, Bayerische Akademie der Wissenschaften, Walther-Meissner-Str. 8, D-85748, Garching, Germany}

\author{C. Putzke}
\affiliation{H. H. Wills Physics Laboratory, University of Bristol, Tyndall Avenue, Bristol, BS8 1TL, United Kingdom}

\author{E. Kampert}
\affiliation{Helmholtz-Zentrum Dresden-Rossendorf, Hochfeld-Magnetlabor Dresden, DE-01314 Dresden, Germany}

\author{M. V. Kartsovnik$^3$}

\author{V. M. Krasnov$^1$}
\email[E-mail: ]{Vladimir.Krasnov@fysik.su.se}

\date{\today }

\begin{abstract}

We experimentally study transport and intrinsic tunneling
characteristics of a single-layer cuprate
Bi$_{2+x}$Sr$_{2-y}$CuO$_{6+\delta}$ with a low superconducting
critical temperature $T_\text{c} \lesssim 4$\,K. It is observed
that the superconducting energy, critical field and fluctuation
temperature range are scaling down with $T_\text{c}$, while the
corresponding pseudogap characteristics have the same order of
magnitude as for high-$T_\text{c}$ cuprates with 20 to 30 times
higher $T_\text{c}$. The observed disparity of the
superconducting and pseudogap scales clearly reveals their
different origins.

\end{abstract}

\maketitle

An interplay between the normal state pseudogap (PG) and
unconventional superconductivity remains one of the highly debated
and controversial issues. For high transition temperature cuprates
the energy scales of the superconducting gap (SG) and the PG are
similar ($\sim 30-50$\,meV). They are close to several oxygen
phonon modes and comparable to the antiferromagnetic exchange
energy. It remains unsettled whether this is a mere coincidence, a
prerequisite for a high $T_\text{c}$, or an indication of the
common origin of the SG and the PG. Within the precursor
superconductivity scenario of the PG, the apparent large
difference between the PG suppression temperature $T^*$ and
$T_\text{c}$ is attributed to large phase fluctuations (for
review, see e.g. Ref.~\cite{Varlamov}). The extent of fluctuations
is determined by the ratio of the superconducting energy gap
$\Delta_\text{SG}$ to the Fermi energy $E_\text{F}$ and the
dimensionality of the system \cite{Varlamov}. Therefore, it is
instructive to compare superconducting and PG characteristics of
homologous series of cuprates with different number of CuO planes
per unit cell because such cuprates have similar Fermi energies
\cite{Kaminski_NP2011,He_Bi2201_Kerr2011,Okada_ARPES2011},
resistivities, anisotropies and a similar dimensionality
\cite{Vedeneev_Hc2,Ando,Watanabe2212}, but exhibit a large
variation of $T_\text{c}$. Since thermal fluctuations decrease
with decreasing temperature and vanish at $T\rightarrow 0$, an
analysis of the interplay between the pseudogap and
superconductivity in very low-$T_\text{c}$ cuprates may provide an
important constraint for theoretical models.

Single crystals of Bi-cuprates represent natural stacks of
atomic-scale intrinsic Josephson junctions
\cite{Kleiner,Katterwe,Yurgens_Bi2201,MQT_Bi2201}. The $c$-axis
transport is caused by interlayer tunneling, which creates the
basis for the intrinsic tunneling spectroscopy technique and
facilitates measurements of bulk electronic spectra of cuprates
\cite{SecondOrder,MR}.

In this work we present combined intrinsic tunneling and transport
measurements of Bi$_{2+x}$Sr$_{2-y}$CuO$_{6+\delta}$
\hbox{(Bi-2201)} single crystals with low $T_\text{c} \lesssim 4$\,K. 
We observe that all superconducting characteristics
(temperature, energy and magnetic field scales) are reduced in
proportion to $T_\text{c}$, but the corresponding pseudogap
characteristics remain similar to that in high-$T_\text{c}$
Bi-2212 and Bi-2223 compounds with 20-30 times larger
$T_\text{c}$. The disparity of the superconducting and
$c$-axis pseudogap scales in Bi-2201 clearly reveals their
different origins.

We study small mesa structures made on top of freshly cleaved
Bi-2201 single crystals. The crystals were grown similar to 
Ref.~\cite{Maljuk}. The $T_\text{c}$ of the crystals depends not only
on oxygen doping $\delta$, but also on the Bi and Sr content and
can be tuned to zero in the stoichiometric $x=y=0$ compound.
As-grown crystals are slightly overdoped (OD) with a $T_\text{c}
\simeq 3.5$\,K. Six or twelve mesas/contacts were fabricated on
each crystal, which allows simultaneous measurements of the
$ab$-plane and the $c$-axis transport characteristics in the
four-probe configuration. In total about ten crystals were studied
and the data shown below are representative for all samples. Most
measurements were done in a flowing gas $^4$He cryostat at 1.8\,K
$<T<$ 300\,K and a magnetic field up to 17\,T, or a $^3$He
cryostat with $T$ down to 270\,mK and a field up to 5\,T.
Complementary measurements were made in pulsed magnetic fields up
to 65\,T. In all cases magnetic field was applied along the
$c$-axis direction.

\begin{figure}[h]
    \includegraphics[width=5.0cm]{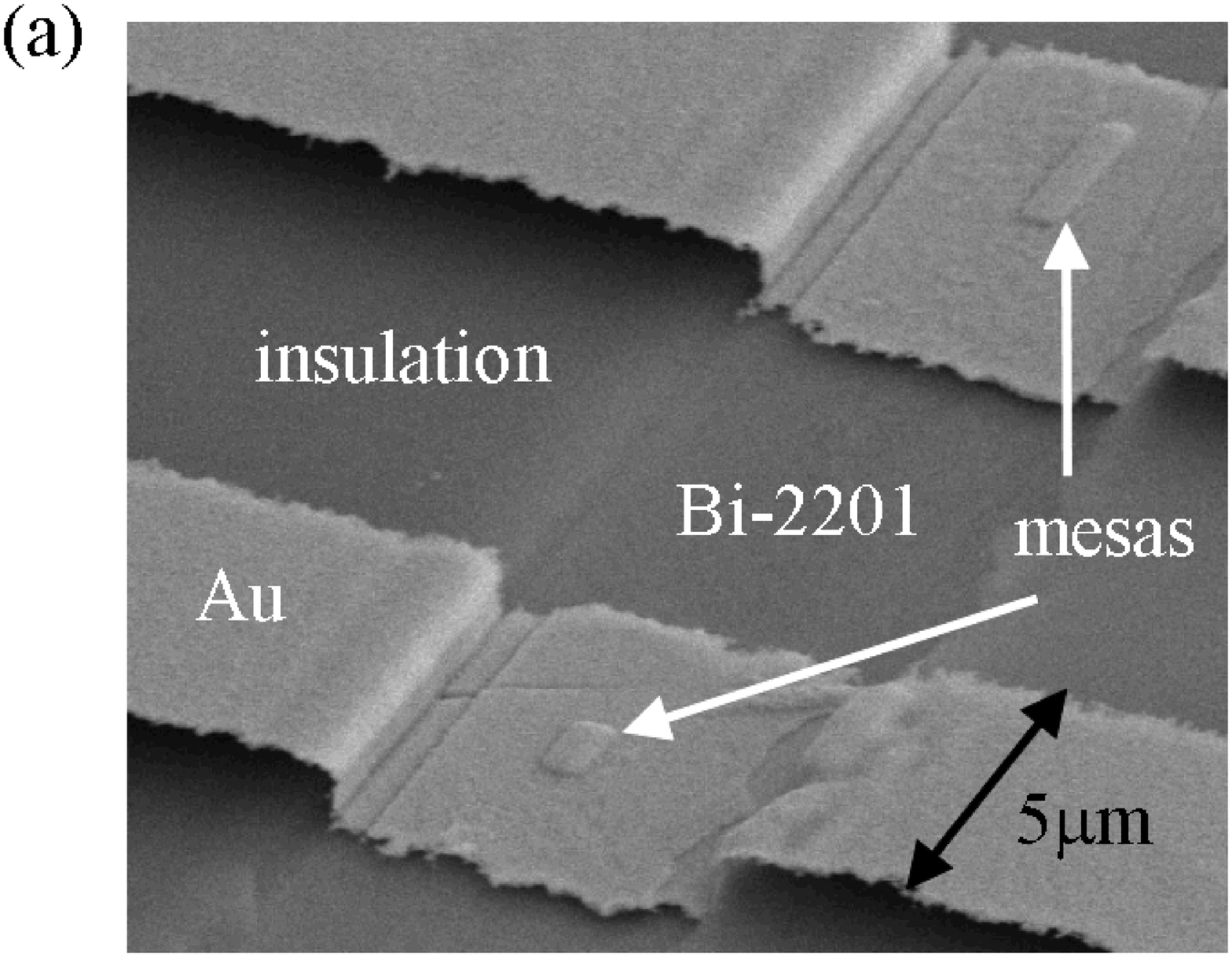}
    \includegraphics[width=6.5cm]{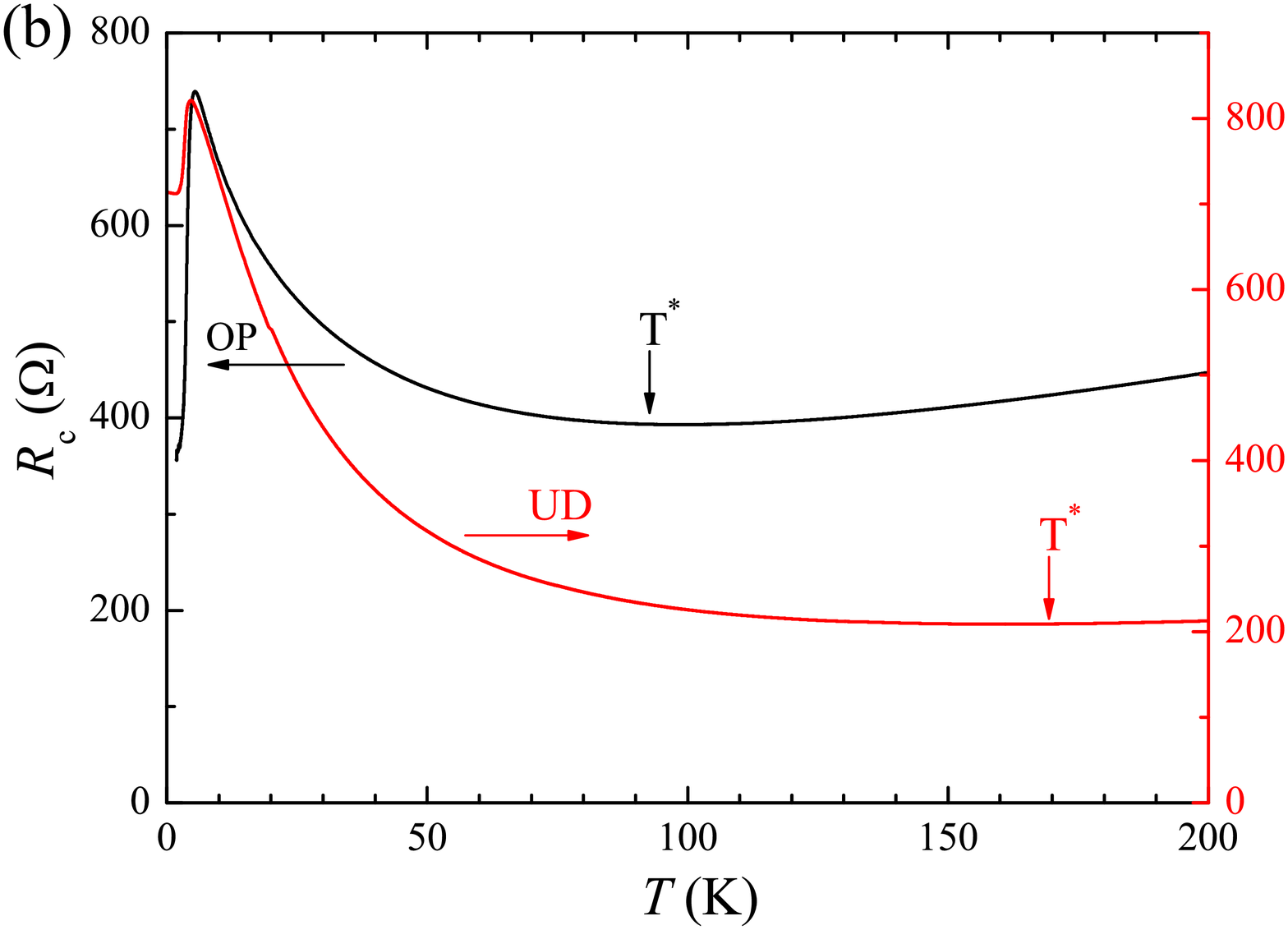}
    \caption{
    (a) SEM image of Bi-2201 mesas. Sub-micron mesas were made
    by direct patterning of Pt masks in a dual-beam SEM/FIB.
    (b) Zero bias c-axis resistance for an optimally doped (OP) mesa
    with $T_\text{c}\simeq4.0$\,K and for an underdoped (UD) mesa
    with $T_\text{c}\simeq3.7$\,K. The pseudogap opening temperatures $T^*$
    are indicated by arrows.}
    \label{fig:fig1}
\end{figure}

In order to minimize self-heating in the intrinsic tunneling
experiments, we limited the height of the mesas to only a few
atomic layers, using slow 250\,eV Ar-ion milling, as well as
reduced lateral mesa sizes down to sub-micrometers, using a
dual-beam SEM/FIB system for mesa pattering or trimming.
Two different miniaturization techniques were used,
providing slightly different mesa properties. In the
first technique we initially made moderately large $\sim 5\times
5\,\upmu$m$^2$ mesas using optical lithography and
subsequently trimmed them to smaller sizes \cite{Katterwe}.
In the second technique we directly patterned
small mesas using electron-beam deposited Pt/C masks, followed by
O-plasma ashing of remaining C. Such mesas are
shown in Fig.~\ref{fig:fig1}\,(a). They are completely covered
by~$\sim 200$\,nm thick gold electrodes, which further improve
the thermal properties of the mesas.

The main difference between both types of mesas is in the size of
the mesa being Ar-milled. Ar-milling leads to a partial
out-diffusion of oxygen. The smaller the mesa area is during
milling, the easier is an in-plane \hbox{O-diffusion}, which leads
to a lower doping of the mesa after milling.
Fig.~\ref{fig:fig1}\,(b) shows the zero-bias $c$-axis resistance
$R_\text{c}(T)$ for mesas made by those two techniques. The mesas
made by the first technique show a metallic behavior at high
temperatures, which turns into a semiconducting-like at $T^*\sim 95$\,K, 
while the mesas made by the second technique show
semiconducting behavior already at $T^* \simeq 170$\,K 
(determined from the minimum in $R(T)$). By
comparing with reported doping dependences of $R_\text{c}(T)$ for
Bi-2201 \cite{Ando,Lavrov}, we conclude that the former mesa is
near optimally doped (OP), while the latter became slightly
underdoped (UD). This is accompanied by a systematic variation of
the critical temperature from $T_\text{c}\simeq 3.5$\,K for as
grown overdoped crystals to \hbox{$T_\text{c} \simeq 4.0$\,K (OP)} 
and $3.7$\,K (UD) for mesas made by the first and second technique,
respectively. The decoupling of  \hbox{$T_\text{c}$ and $T^*$} in our
low-$T_\text{c}$ Bi-2201 allows a clear analysis of the pure PG behavior
above $T_\text{c}$ without an interference from superconductivity. From
Fig.~\ref{fig:fig1}\,(b) it is clear that the pseudogap $T^*$
grows rapidly with underdoping, similar to that for La-doped
\hbox{Bi-2201} \hbox{(Bi(La)-2201)} \cite{Lavrov,Yurgens_Bi2201}
and consistent with the existence of a quantum critical doping
point \cite{Balakirev}.

\begin{figure*}[t]
    \includegraphics[width=\textwidth]{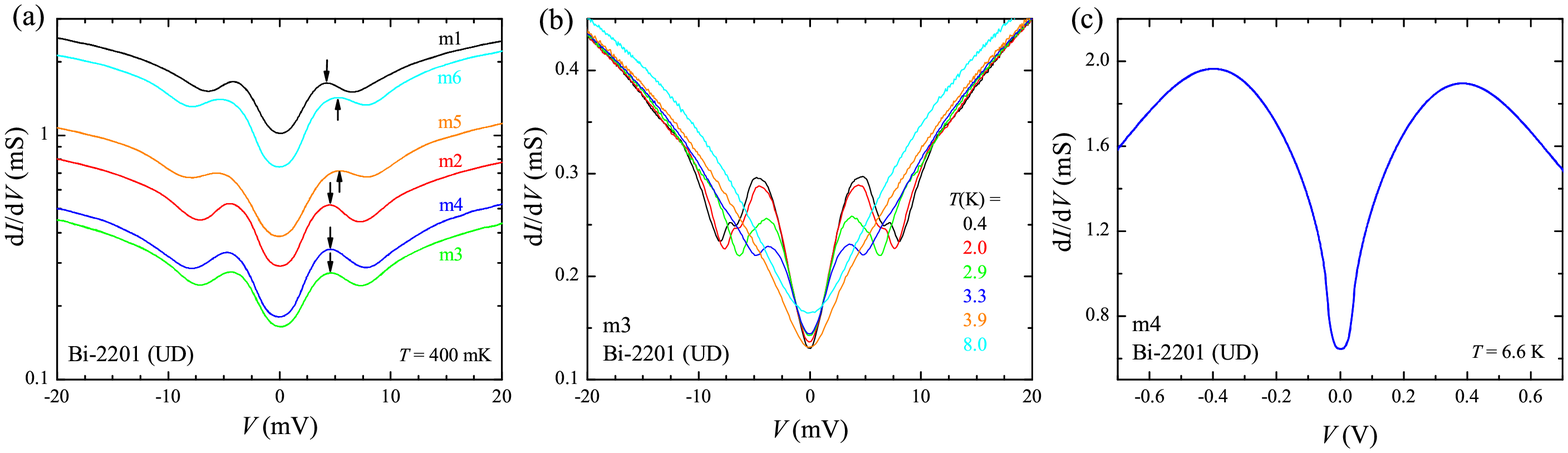}
    \caption{(Color online)
    (a) Intrinsic tunneling conductances (in the semi-log scale) at 400\,mK
    for UD mesas with different areas (0.8 to 2.5\,$\upmu$m$^2$), 
    fabricated on the same crystal. Downward and upward arrows mark 
    sum-gap peaks for mesas with \hbox{$N=5$ and $6$} junctions, respectively. It is seen that the
    $\mathrm dI/\mathrm dV(V)$ characteristics retain the same shape with a
    peak and a dip, despite a significant difference in mesa sizes and
    power dissipation, which indicates the lack of distortion by self-heating.
    (b) Temperature dependence of $\mathrm dI/\mathrm dV(V)$ curves
    for the smallest mesa. It is seen that both the peak and
    dip features disappear in the BCS mean-field manner at
    $T\rightarrow T_\text{c}$.
    (c) A large bias $\mathrm dI/\mathrm dV(V)$
    at 6.6\,K showing the pronounced conductance suppression
    caused by the $c$-axis pseudogap at $T>T_\text{c}$. Note the large scale of
    the pseudogap hump, which corresponds to $\Delta_{PG} \simeq 40$\,meV.}
    \label{fig:dIdV}
\end{figure*}

An unambiguous discrimination of spectroscopic features from
self-heating artifacts can be obtained by studying the size
dependence of current-voltage \hbox{($I$-$V$)} characteristics:
spectroscopic features are material properties and should be size
independent, while heating strongly depends on the mesa size
\cite{Heat2005,SecondOrder}. Fig.~\ref{fig:dIdV}\,(a) shows
intrinsic tunneling conductances $\mathrm dI/\mathrm dV(V)$ 
at low $T$ and $H$ for UD mesas with different areas (from
0.8 to 2.5\,$\upmu$m$^2$). They are fabricated on the same crystal  
and the mesa conductance scales with their area. It can be seen that all
curves maintain similar shapes in the semi-logarithmic scale,
despite significant differences in dissipation powers (the power
at the peaks is ranging between $\approx1.2$ and $7.4$\,$\upmu$W).
The size-independence indicates that there is no significant
distortion by self-heating.

From Fig.~\ref{fig:dIdV}\,(a) it can be seen that the intrinsic
tunneling characteristics exhibit a peak-dip structure, typical
for tunneling characteristics of (underdoped) cuprates
\cite{MiyakawaBi2201,Fisher2223,Hudson_CDWinBi2201,Kato_STM2010,Kugler_Bi2201STM,Doping}.
The peak is attributed to the superconducting sum-gap singularity
at $V_\text{peak}=2\Delta_\text{SG}N/e$, where $N$ is the number
of stacked intrinsic junctions in the mesa \cite{SecondOrder,MR}.
The dip is usually ascribed to an interaction with a collective
bosonic mode with energy $\Omega_\text{B}$, in which case
$eV_\text{dip}=2\Delta_\text{SG} + \Omega_\text{B}$ (per junction)
\cite{MiyakawaBi2201,Fisher2223}. The ratio
$\Omega_\text{B}/2\Delta_\text{SG} \simeq 0.62$ can be extracted
explicitly from the $\mathrm dI/\mathrm dV(V)$ characteristic in
Fig.~\ref{fig:dIdV}\,(b). It is very close to the average value
$\simeq 0.64$, deduced from inelastic neutron scattering data for
different cuprates \cite{Yu2009}.

In order to evaluate absolute values of $\Delta_\text{SG}$ and
$\Omega_\text{B}$, we have to estimate the number of junctions $N$
in the mesas. For high-$T_\text{c}$ Bi-2212
\cite{Kleiner,SecondOrder} and \hbox{(Bi(La)-2201)}
\cite{Yurgens_Bi2201,MQT_Bi2201} this is simply done by counting
quasiparticle branches in the $I$-$V$ characteristics. However, in
low-$T_\text{c}$ Bi-2201, the branches are not distinguishable
\cite{Yurgens_Bi2201}. Still it is possible to estimate $N$ by
comparing the peak voltages of mesas with different heights. Since the
uniformity of Ar-milling is typically about one crystallographic unit cell,
there is a certain probability that mesas on the same
crystal have either $N$ or $N+1$ junctions. In
Fig.~\ref{fig:dIdV}\,(a) this is seen as two series of peaks with
slightly different voltages, marked by downward and upward
arrows, respectively. Assuming that the ratio of voltages is
$(N+1)/N$, we obtain that one series of mesas contains $N=5$ and
the other $N=6$ intrinsic junctions. This yields 
$\Delta_\text{SC}\simeq 0.55 \pm 0.1$\,meV and 
$2\Delta/k_\text{B} T_\text{c}\simeq 3.5 \pm 0.6$, 
which is typical for weak-coupling superconductors and consistent 
with previous studies on Bi(La)-2201 with higher $T_\text{c}$
\cite{Yurgens_Bi2201,Ideta_ARPES,Kato_STM2010,MiyakawaBi2201}. The
corresponding bosonic mode energy is also small, 
$\Omega_\text{B}\simeq 0.7$\,meV, and is clearly not related to the 
much higher phonon or antiferromagnetic energies. This observation
demonstrates that $\Omega_\text{B}$ is determined solely by
$\Delta_\text{SG}(T)$ and supports the interpretation of the mode
as an exciton of Bogoliubov quasiparticles caused by a 
\hbox{$d$-wave} symmetry of the order parameter \cite{Chubukov}.

Fig.~\ref{fig:dIdV}\,(b) shows the $T$-dependence of 
$\mathrm dI/\mathrm dV(V)$ characteristics for the smallest UD mesa 
at $H=0$\,T. One can clearly see that both, the peak and the dip
decrease in amplitude, move to lower voltages with increasing $T$
and vanish at $T_\text{c} \simeq 3.7$\,K in the BCS mean-field
manner (flat at $T \ll T_\text{c}$ and a fast decay at $T\rightarrow T_\text{c}$).
Above $T_\text{c}$, the $\mathrm dI/\mathrm dV(V)$ remains
V-shaped due to the presence of the $c$-axis PG, which leads to
thermal activation behavior of interlayer transport
\cite{SecondOrder,KatterwePRL2008}. Fig.~\ref{fig:dIdV}\,(c) shows
a large bias $\mathrm dI/\mathrm dV(V)$ of an UD mesa at
$T=6.6$\,K $>T_\text{c}$. The pseudogap hump is clearly visible at
$V \simeq 0.4$\,V, corresponding to 
$\Delta_\text{PG} \simeq40$\,meV. It is remarkably similar to that 
for Bi-2212 \cite{Doping,Yurgens_Bi2201,KatterwePRL2008,SecondOrder}, 
despite a more than 20 times smaller $T_\text{c}$ and $\Delta_\text{SG}$.
The large $\Delta_\text{PG}$ leads to the high onset temperature
$T^* \sim 170$\,K in the UD mesa, see Fig.~\ref{fig:fig1}\,(b).
Obviously, there is a large difference between the superconducting
and the $c$-axis pseudogap energies in low-$T_\text{c}$ Bi-2201.

From Fig.~\ref{fig:dIdV}\,(b) it is clear that the
thermal-activation-like \cite{KatterwePRL2008,SecondOrder},
V-shaped PG characteristics coexists with superconductivity at
$T<T_\text{c}$. This makes it difficult to separate
superconducting and pseudogap-related phenomena below $T_\text{c}$ by
looking solely on the $T$-variation of experimental
characteristics \cite{Kaminski_NP2011}. A more unambiguous
discrimination between them can be done by analyzing magnetic
field effects \cite{MR}. The upper critical field $H_\text{c2}$,
required to suppress spin-singlet superconductivity, is limited by
the paramagnetic (Zeeman) effect. This yields 
$\mathrm dH_\text{c2}/\mathrm dT(T=T_\text{c}) = -2.25$\,T/K 
for weak coupling d-wave superconductors and lower values for strong
coupling cases \cite{ParamagneticLimit}. Therefore, $H_\text{c2}$
for low-$T_\text{c}$ Bi-2201 should be much lower and easier
accessible than for \hbox{high-$T_\text{c}$} \hbox{Bi-2212}
with $H_\text{c2} \sim 100$\,T \cite{MR}.

\begin{figure*}[t]
    \includegraphics[width=\textwidth]{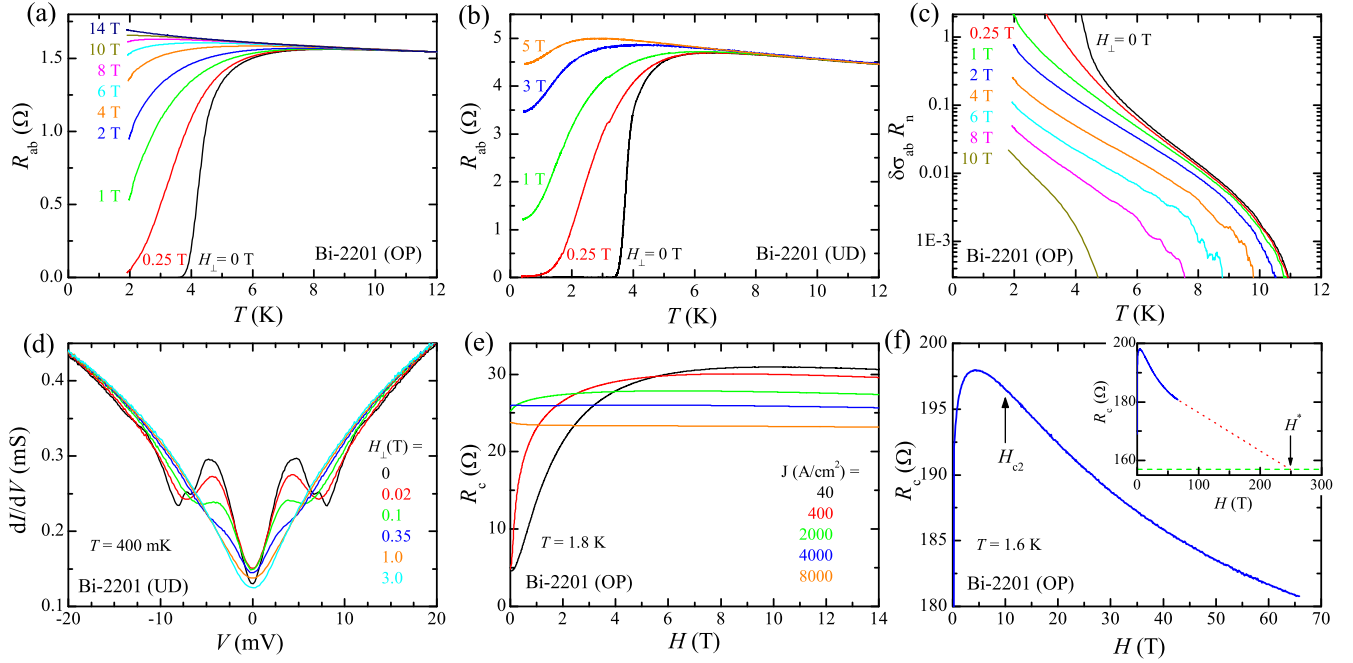}
    \caption{(Color online) Zero-bias $ab$-plane resistance at
    different $c$-axis magnetic fields for (a) an OP mesa and (b) an
    UD mesa. A complete suppression of $T_\text{c}$ at $H_{c2}\simeq 10$\,T is seen.
    (c) Normalized in-plane fluctuation (excess) conductance
    \hbox{$[1/R_\text{ab}(T,H)-1/R_\text{n}(T)]R_\text{n}(T)$ vs. $T$} for
    the data from (a). It is seen that fluctuations decay rapidly
    (almost exponentially) with increasing $T$ and that the region of
    significant fluctuations expand to about twice the $T_\text{c}$.
    (d) Magnetic field dependence of intrinsic tunneling characteristics
    for an UD mesa at 400\,mK. The sum-gap peak and the dip are washed
    away in a $c$-axis field of a few Tesla. (e) Bias current dependence
    of the $c$-axis tunneling magnetoresistance (MR) for an OP mesa.
    The MR is large at small current densities but vanishes at large
    bias, corresponding to the nearly ohmic normal (tunnel) resistance
    part of the $I$-$V$.
    (f) $c$-axis MR in a pulsed field. A profound negative MR at
    large fields is attributed to suppression of the pseudogap.
    The inset demonstrates an estimation of the PG
    suppression field $H^*\sim 250\pm 50$\,T by extrapolation to a large
    bias, $H$- and \hbox{$T$-independent} tunneling resistance (dashed
    horizontal line).}
    \label{fig:field}
\end{figure*}

Fig.~\ref{fig:field}\,(a) shows zero-bias $ab$-plane resistances,
$R_\text{ab}(T)$, at different magnetic fields in the $c$-axis
direction for the OP mesa with $T_\text{c}(H=0) = 4.0 \pm 0.5$\,K.
With increasing~$H$, $T_\text{c}$ decreases and is no longer
visible at $H > 10$\,T. Fig.~\ref{fig:field}\,(b) shows similar
data for the UD mesa, measured down to sub-Kelvin temperatures.
The graph demonstrates that the $T_\text{c}$ is indeed completely
suppressed by a field of a few Tesla.

A significant positive magnetoresistance can be seen in the
$R_\text{ab}(T,H)$ curves in Figs.~\ref{fig:field}\,(a) and (b) which
persists well above $T_\text{c}$, indicating a broad
superconducting fluctuation region. Fig.~\ref{fig:field}\,(c)
represents the normalized excess conductance
\hbox{$[1/R_\text{ab}(T,H)-1/R_\text{n}(T)]R_\text{n}(T)$ vs. $T$}
for the data from Fig.~\ref{fig:field}\,(a). Here we used the
curve at $H=14$\,T as normal state resistance 
\hbox{$R_\text{n}(T)=R_\text{ab}(T, H=14\,\text{T})$}. 
The fluctuation MR of \hbox{Bi-2201} is decreasing
almost exponentially with increasing $T$, which was also observed
for Bi-2212 \cite{MR}. The MR falls to less than $1\%$ at a
relative temperature \hbox{$T/T_\text{c}(H) \simeq 2.3$}, comparable to
that for Bi(La)-2201 with higher $T_\text{c}$ \cite{Lavrov}. Even though
we can not exclude an additional influence of sample inhomogeneity
on the broadening of the superconducting transition, it is clear
that the temperature range of superconducting fluctuations is
determined by $T_\text{c}$ and not by the much larger 
\hbox{$T^*\sim 100-200$\,K.}

Fig.~\ref{fig:field}\,(d) shows the interlayer tunneling
conductance $\mathrm dI/\mathrm dV(V)$ at different $c$-axis
magnetic fields and at \hbox{$T=400$\,mK} for a small UD mesa.
Both, the peak and the dip are moving to lower voltages and smear
out in a correlated manner with increasing $H$. The tunneling MR
remains significant at a bias several times larger than at the
sum-gap peak, corresponding to a dissipation power about an order
of magnitude larger than at the peak. This, again, indicates the
absence of significant self-heating at the peak. From the combined
transport, Fig.~\ref{fig:field}\,(a), and magnetotunneling
measurements, Fig.~\ref{fig:field}\,(d), we estimate
$H_\text{c2}(0)=10 \pm 2$\,T and $H_\text{c2}(0)/T_\text{c} = 2.5\pm 1$\,T/K, 
consistent with previous reports \cite{NMR_Bi2201,Vedeneev_Hc2,MR}. 
This value is close to the paramagnetic limit, which rules out a 
major underestimation of $H_\text{c2}$.

In Fig.~\ref{fig:field}\,(e) the bias dependence of the interlayer 
tunneling MR for a moderately large OP mesa at \hbox{$T=1.8$\,K} is shown. 
It demonstrates that the MR has a strong bias dependence and
vanishes at large bias, corresponding to the normal (tunnel)
resistance part of the $I$-$V$, which is the consequence of state
conservation \cite{MR}.

Fig.~\ref{fig:field}\,(f) shows the zero-bias $c$-axis resistance
$R_\text{c}(H)$ of a smaller OP mesa from the same sample, 
measured in a pulsed magnetic field
up to 65\,T. Two contributions to the MR are clearly seen: At low
fields there is a positive MR due to suppression of the
interlayer supercurrent, as in panel (e), and at higher fields a
transition to a profound negative MR occurs, which does not
saturate at $65$\,T. A very large characteristic field, 
$H^*\sim 250 \pm 50$\,T, is estimated by linear
extrapolation to an (almost) $H$- and $T$- independent,
high bias tunnel resistance (dashed horizontal line in the inset of
Fig.~\ref{fig:field}\,(f)). Even though such behavior has been
reported before \cite{Ando,Vedeneev_Hc2,Shibauchi}, its
interpretation for high-$T_\text{c}$ cuprates is complicated by a
large $H_\text{c2}\sim 100$\,T \cite{MR}. The low-$T_\text{c}$ of
our Bi-2201 leads to a complete separation of the superconducting
$H_\text{c2}\sim 10$\,T and the pseudogap $H^*\sim 250$\,T
field scales. We attribute the negative $c$-axis magnetoresistance
at $H^*\gg H_\text{c2}$ to a field suppression of the PG,
associated either with an (antiferro-) magnetic
\cite{He_Bi2201_Kerr2011,NMR_Bi2201} or spin/charge density wave
order \cite{Hudson_CDWinBi2201}.

To summarize, combined transport and intrinsic tunneling
measurements where performed on a \hbox{low-$T_\text{c}$} \hbox{Bi-2201}
cuprate. Despite the very low \hbox{$T_\text{c} \lesssim 4$\,K}, the
\hbox{$c$-axis} pseudogap characteristics 
(\hbox{$\Delta_\text{PG}\simeq 40$\,meV}, 
\hbox{$T^*\sim 100-300$\,K}, and \hbox{$H^* \sim 200-300$\,T}) are
remarkably similar to those for high-$T_\text{c}$ Bi-2212 and Bi-2223
compounds with 20-30 times higher $T_\text{c}$. On the contrary, all
superconducting properties
(\hbox{$2\Delta_\text{SG}/k_\text{B}T_\text{c} \simeq 3.5 \pm0.6$}, 
the bosonic mode 
\hbox{$\Omega_\text{B}/k_\text{B}T_\text{c}\simeq 2.2$}, 
\hbox{$H_\text{c2}(0)/T_\text{c} \simeq 2.5 \pm 1$\,T/K}) 
and the fluctuation region are scaling down with $T_\text{c}$. 
The striking difference between the energy, temperature and 
magnetic-field scales, associated with superconductivity, and the 
pseudogap clearly reveals their different origins. 
Note that the observed scaling of superconducting properties with
$T_\text{c}$ is similar to that for high-$T_\text{c}$ cuprates
\cite{Ideta_ARPES,Yu2009,Vedeneev_Hc2,Kato_STM2010,Lavrov,Dubroka_2011}.
Such a universality within a homologous family of Bi-based
cuprates with similar carrier concentrations, resistivities,
anisotropies and layeredness, but largely different $T_\text{c}$
suggests that the superconducting transition maintains the same
character with a similar relative strength of superconducting
fluctuations, irrespective of $T_\text{c}$.

Support by the K.\:\&\:A.\:Wallenberg foundation, the Swedish
Research Council, and the SU-Core facility in Nanotechnology is
gratefully acknowledged. The high-field experiment was done with
support from EuroMagNET II under the EC Contract No. 228043.
A. Maljuk thanks C. T. Lin (MPI-Stuttgart) for helpful discussions 
regarding the crystal growth.

\end{document}